\pdfoutput=1
\documentclass[twocolumn,showpacs,preprintnumbers,aps,prl,superscriptaddress]{revtex4}%
\usepackage{graphicx}
\usepackage{amsmath}
\usepackage{amsfonts}
\usepackage{amssymb}
\usepackage{hyperref}
\begin{document}

\title{Broad Feshbach resonance in the $^{6}$Li-$^{40}$K mixture}
\author{T.G. Tiecke}
\affiliation{Van der Waals-Zeeman Institute of the University of
Amsterdam, Valckenierstraat 65, 1018 XE, The Netherlands}
\author{M.R. Goosen}
\affiliation{Eindhoven University of Technology, P.O. Box 513,
5600 MB Eindhoven, The Netherlands}
\author{A. Ludewig}
\author{S.D. Gensemer}
\altaffiliation{Present address: Ethel Walker School, 230 Bushy
Hill Rd, Simsbury, CT 06070, United States.}
\author{S. Kraft}
\altaffiliation{Present address: Physikalisch-Technische
Bundesanstalt, Bundesallee 100, D-38116 Braunschweig.}
\affiliation{Van der Waals-Zeeman Institute of the University of
Amsterdam, Valckenierstraat 65, 1018 XE, The Netherlands}
\author{S.J.J.M.F. Kokkelmans }
\affiliation{Eindhoven University of Technology, P.O. Box 513,
5600 MB Eindhoven, The Netherlands}
\author{J.T.M. Walraven}
\affiliation{Van der Waals-Zeeman Institute of the University of
Amsterdam, Valckenierstraat 65, 1018 XE, The Netherlands}
\date{\today}

\begin{abstract}
We study the widths of interspecies Feshbach resonances in a
mixture of the fermionic quantum gases $^{6}$Li and $^{40}$K. We
develop a model to calculate the width and position of all
available Feshbach resonances for a
system. Using the model we select the optimal resonance to study the $^{6}$%
Li/$^{40}$K mixture. Experimentally, we obtain the asymmetric Fano
lineshape of the interspecies elastic cross section by measuring
the distillation rate of $^{6}$Li atoms from a potassium-rich
$^{6}$Li/$^{40}$K mixture as a function of magnetic field. This
provides us with the first experimental
determination of the width of a resonance in this mixture, $\Delta B=1.5(5)~%
\mathrm{G}$. Our results offer good perspectives for the
observation of universal crossover physics using this
mass-imbalanced fermionic mixture.
\end{abstract}

\maketitle

A decade of experiments with degenerate fermionic quantum gases
has delivered major scientific advances as well as a whole new
class of quantum many-body systems
\cite{jin98,giorgini08,varenna08}. Feshbach resonances
\cite{verhaar93} played a central role in this development as they
offer exceptional control over the interatomic interactions at
low-temperatures \cite{chin08}. In gases with the appropriate spin
mixture the sign and magnitude of the $s$-wave scattering length
$a$ can be tuned to any positive or negative value by choosing the
proper magnetic field in the vicinity of a resonance. In the case
of fermionic atoms the role of Feshbach resonances is especially
remarkable because Pauli exclusion dramatically suppresses
three-body losses to deeply bound molecular states \cite%
{fermiMolecules03,petrov04}. The tunability has been used with
great success in two-component Fermi gases of $^{6}$Li and of
$^{40}$K to study and control pairing mechanisms, both of the
Cooper type on the attractive side of the resonance $\left(
a<0\right) $ \cite{BCSpairing} and of the molecular type on the
repulsive side $\left( a>0\right) $ \cite{molBECs}. In particular
the universal crossover from the superfluidity of a molecular
Bose-Einstein condensate (BEC) towards the Bardeen, Cooper,
Schrieffer (BCS) limit received a lot of attention \cite{BECBCS}.
Essential for these studies
is the availability of sufficiently broad Feshbach resonances in the $^{6}$%
Li and $^{40}$K homonuclear gases.

Recently the study of heteronuclear fermionic mixtures has
strongly gained in interest due to its additional mass imbalance.
Theoretical studies on these mixtures include e.g.: superfluidity
\cite{baranov08}, phase separation \cite{bausmerth09}, crystalline
phases \cite{dimaWigner07},
exotic pairing mechanisms \cite{wilczek05} and long-lived trimers \cite%
{petrov09}. Many of these studies require the mixture to be
strongly interacting and in the universal limit, i.e. the
scattering length should be very large and the only parameter that
determines the two-body interaction. 
Recently, the first mass-imbalanced ultracold fermionic mixture
has been
realized, namely, a mixture of the only stable fermionic alkaline species $%
^{6}$Li and $^{40}$K \cite{taglieber08}. The basic interaction
properties of the $^{6}$Li/$^{40}$K system were established in
experiments by the Innsbruck group \cite{wille08}, in which the
loss features of $13$ Feshbach resonances were observed and
assigned. The first $^{6}$Li$^{40}$K molecules were recently
reported from Munich \cite{voigt09}. Despite this experimental
progress a sufficiently broad Feshbach resonance to use the $^{6}$Li/$^{40}$%
K system for universal studies has not been reported so far.

In this Letter we identify and characterize the optimal Feshbach
resonance of the $^{6}$Li/$^{40}$K mixture. We develop a generic
model to estimate the positions and widths of all available
$s$-wave Feshbach resonances in quantum gases. By applying this
model to the two-component $^{6}$Li/$^{40}$K mixtures stable
against spin exchange we select the optimal resonance compromising
between resonance width and convenience for detection. We present
the first measurement of a resonance width in the
$^{6}$Li/$^{40}$K mixture by measuring the asymmetric lineshape
(Fano profile) of the inter-species elastic cross section near the
Feshbach resonance.
We measure the rate of distillation of $^{6}$Li atoms from a potassium-rich $%
^{6}$Li/$^{40}$K mixture confined in an optical dipole trap.
The measured resonance width is shown to be promising for reaching
the universal regime in the $^{6}$Li/$^{40}$K mixture.

In search for broad and accessible Feshbach resonances we extend
the Asymptotic Bound-state Model (ABM) \cite{wille08} to include
the description of resonance widths. We start from the two-body
Hamiltonian for the relative motion
\begin{equation}
\mathcal{H}=\mathbf{p}^{2}/2\mu +\mathcal{V}+\mathcal{H}^{int}=\mathcal{H}%
^{rel}+\mathcal{H}^{int},  \label{1}
\end{equation}%
containing the relative kinetic energy with $\mu $ the reduced
mass, the electron spin dependent central interatomic interaction
$\mathcal{V}$, and
the internal energy $\mathcal{H}^{int}$ of the two atoms. Here we restrict $%
\mathcal{H}^{int}$ to the hyperfine and Zeeman terms and consider
$s$-wave interactions only. Instead of solving coupled radial
Schr\"{o}dinger equations, the ABM approach relies on knowledge of
the binding energies of the highest bound states in the two-body
system. This is sufficient to determine the scattering properties,
and in particular the position of
Feshbach resonances. For $^{6}$Li/$^{40}$K only the least bound levels of $%
\mathcal{H}^{rel}$ are relevant and can be obtained using the eigenvalues $%
E_{S}$ of the least bound states in the electron-spin singlet
($S=0$) and
triplet ($S=1$) potentials as free parameters; here we adapt $E_{0}$ and $%
E_{1}$ from Ref.~\cite{wille08}.

The mixture is prepared in one of the two-body hyperfine eigenstates of $%
\mathcal{H}^{int}$ at magnetic field $B$, referred to as the
$P$-channel or
open channel; denoted via the $B=0$ hyperfine quantum numbers as $%
|f,m_{f}\rangle _{\alpha }\otimes |f,m_{f}\rangle _{\beta }$. The
corresponding energy of two free atoms at rest defines a
$B$-dependent
reference value representing the threshold between the scattering states ($%
E>0$) and the bound states ($E<0$) of $\mathcal{H}$. We define
$\mathcal{H}$ relative to this threshold energy. A complete basis
for the spin properties is defined via the quantum numbers $S$,
its projection $M_{S}$, and the projection of the nuclear spins
$\mu _{\alpha }$ and $\mu _{\beta }$, while requiring that the
total projection $M_{S}+\mu _{\alpha }+\mu _{\beta
}=m_{f_{\alpha }}+m_{f_{\beta }}=M_{F}$ is fixed. By diagonalizing $\mathcal{%
H}$ starting from this `singlet-triplet' basis we find the
energies of the bound states, and the Feshbach resonances are
localized at the magnetic fields where they intersect with the
energy of the threshold.

Threshold effects cause the approximately linear magnetic field
dependence of the bound-state energies to change to quadratic
behavior close to the field of resonance \cite{varenna08,chin08}.
This provides information about the width of a Feshbach resonance.
The ABM, as discussed thus far, does not show these threshold
effects, which is not surprising because the threshold is not
explicitly build into the theory; it is merely added as a
reference value for comparison with the ABM eigenvalues.

However, the ABM contains all ingredients to obtain also the
resonance width. The Hamiltonian (\ref{1}) describes all two-body
bound states, belonging to both open and closed channels. The
width depends on the coupling between the open channel and the
various closed channels, which is determined after two basis
transformations to identify the open channel and the resonant
closed channel respectively. First, we separate the open
channel $P$, as defined above, from all other channels: the closed channels $%
Q$ \cite{moerdijk95}. This is realized with a basis transformation
from the singlet-triplet basis to the $|f,m_{f}\rangle _{\alpha
}\otimes |f,m_{f}\rangle _{\beta }$ basis. In this basis we
identify the open channel, namely the hyperfine state in which the
system is experimentally prepared. We refer to this diagonal
subspace as $\mathcal{H}_{PP}$, a single matrix element that we
identify with the (bare) open-channel bound-state energy $\epsilon
_{P}=-\hbar ^{2}\kappa _{P}^{2}/2\mu $. Second, we perform
a basis transformation that diagonalizes the closed-channel subspace $%
\mathcal{H}_{QQ}$, leaving the open-channel $\mathcal{H}_{PP}$
unaffected. The $\mathcal{H}_{QQ}$ matrix contains the
closed-channel bound-state energies $\epsilon _{Q}$ disregarding
the coupling to $\mathcal{H}_{PP}$. This transformation allows to
identify the resonant closed channel. The
coupling between the open and the resonant closed channel is referred to as $%
\mathcal{H}_{PQ}$ and is a measure for the resonance width.

To obtain the magnetic field width of the resonance from
$\mathcal{H}_{PQ}$ we use Feshbach's resonance
theory~\cite{feshbach,marcelis}: a closed-channel bound state
acquires a finite width $\Gamma $ and its energy
undergoes a shift $\Delta _{res}$. If the binding energy of a certain $Q$%
-channel bound state $|\phi _{Q}\rangle $ is sufficiently close to
the threshold, we can effectively consider a two-channel problem
where the
complex energy shift is given by~\cite{marcelis}%
\begin{equation}
\mathcal{A}(E)=\Delta _{res}(E)-\frac{i}{2}\Gamma (E)=\frac{\mu }{\hbar ^{2}}%
\frac{-i\mathrm{A}}{\kappa _{P}(k-i\kappa _{P})},  \label{2}
\end{equation}%
where $\mathrm{A}=|\langle \phi _{P}|\mathcal{H}_{PQ}|\phi
_{Q}\rangle |^{2}$
is the coupling strength to the $P$-channel bound state $|\phi _{P}\rangle $%
. For $k\rightarrow 0$ the expression $\Gamma \equiv \hbar
^{2}k/\mu R^{\ast }$ defines the characteristic length $R^{\ast
}=\hbar^{2}/(2 \mu a_{bg}\mu _{rel}\Delta B)$ of the resonance
\cite{petrov04a}, where $\mu _{rel}=\partial \epsilon
_{Q}/\partial B|_{B=B_{0}}$ is the magnetic moment of the bare $Q$
channel relative to the open channel threshold. The binding energy
$E=\hbar ^{2}k^{2}/2\mu $ of the dressed bound state is obtained
by solving the pole equation of the scattering matrix, given by
$E-\epsilon _{Q}-\mathcal{A}(E)=0$, assuming that near threshold
the bare bound state can be approximated by $\epsilon _{Q} =
\mu_{rel} (B-B_0)-\Delta _{res}$. Close to threshold we obtain for
the dressed bound state energy
\begin{equation}
E=-\left(\frac{2 |\epsilon_P|^{3/2}}{A} \mu_{rel}(B-B_0) \right)^2
\label{3}
\end{equation}%
retrieving the characteristic quadratic dependence of the
molecular state on the magnetic field. Using the dispersive
formula for the field dependence of the scattering length near a
Feshbach resonance, $a(B)=a_{bg}\left( 1-\Delta
B/(B-B_{0})\right)$, we obtain an expression for the magnetic field width $%
\Delta B$ of the resonance
\begin{equation}
\mu _{rel}\Delta B=\frac{a^{P}}{a_{bg}}\frac{\mathrm{A}}{2
|\epsilon _{P}|}. \label{4}
\end{equation}%
The off-resonance scattering is described by the background
scattering
length $a_{bg}=a_{bg}^{P}+a^{P}$, where $a_{bg}^{P}\approx r_{0}$ and $%
a^{P}=\kappa _{P}^{-1}$. Here $r_{0}\equiv (\mu C_{6}/8\hbar
^{2})^{1/4}\simeq 41\,a_{0}$ is the inter-species Van der Waals range, with $%
C_{6}$ the Van der Waals coefficient and $a_{0}$ the Bohr radius.

\begin{figure}
[ptb]
\begin{center}
\includegraphics[
trim=0.200957in 0.222611in 0.408181in 0.448353in,
natheight=3.478300in, natwidth=4.177900in, height=6.3856cm,
width=8.1012cm
]%
{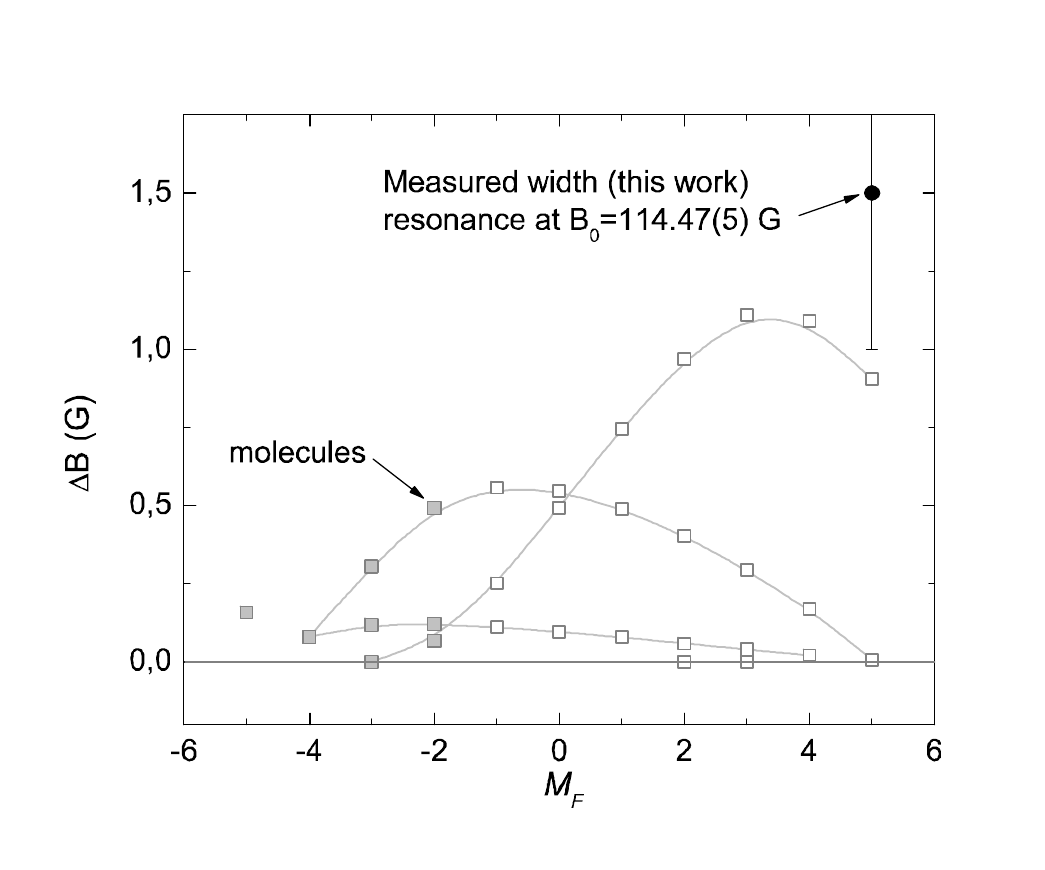}%
\caption{ABM calculated widths of all $s$-wave Feshbach resonances
in stable two-component $^{6}$Li/$^{40}$K mixtures below
$500~\mathrm{G}$. The lines are a guide to the eye. The point at
$M_{F}=-5$ corresponds to the
$|1/2,-1/2\rangle_{\mathrm{Li}}-|9/2,-9/2\rangle_{\mathrm{K}}$
mixture. All other mixtures contain the $^{6}$Li ground state
$|1/2,+1/2\rangle _{\mathrm{Li}}$. Solid black dot: width
measurement reported in this work. The mixtures with
$-M_{F}=5,4,3,2$ (gray squares) where studied in Ref.\thinspace
\cite{wille08}. The resonance used in Ref.~\cite{voigt09} for
molecule
formation is indicated with an arrow.}%
\label{fig.:compilation}%
\end{center}
\end{figure}

The results for all $s$-wave resonances in two-component
$^{6}$Li/$^{40}$K mixtures stable against spin exchange below
$500~\mathrm{G}$ are shown in Fig.~\ref{fig.:compilation}. The
widest resonances for the $^{6}$Li/$^{40}$K mixture are found to
be of the order of $1$~Gauss. From these results the
optimal resonance is selected to be the resonance in the $M_{F}=5$ manifold $%
|1/2,+1/2\rangle _{\mathrm{Li}}\otimes |9/2,+9/2\rangle
_{\mathrm{K}}$ with the predicted position of
$B_{0}=114.7~\mathrm{G}$ as obtained with the ABM
parameters $E_{0,1}$ from Ref.~\cite{wille08}. The predicted width is $%
\Delta B=0.9~\mathrm{G}$. This value is known to slightly
underestimate the actual width \cite{tieckeTBP,tiemannPC}. The
resonance in the $M_{F}=3$
manifold, $|1/2,+1/2\rangle _{\mathrm{Li}}\otimes |9/2,+5/2\rangle _{\mathrm{%
K}}$ is predicted to be the broadest, $20\%$ wider than the
$M_{F}=5$ resonance. However, because the $|9/2,+9/2\rangle
_{\mathrm{K}}$ state has an optical cycling transition,
facilitating detection in high magnetic field, the $M_{F}=5$
resonance is favorable for experimental use. Therefore, this
resonance offers the best compromise between resonance width and
an experimentally favorable internal state.

Our procedure to create an ultracold mixture of $^{6}$Li and
$^{40}$K is described in detail elsewhere
\cite{tiecke09b,tiecke09}. Here we briefly summarize the
procedure. We perform forced evaporative cooling on both species
in an optically plugged magnetic quadrupole trap \cite{davis95}. A
small amount of spin-polarized $^{6}$Li in the $|3/2,+3/2\rangle _{\mathrm{Li%
}} $ hyperfine state is sympathetically cooled by rethermalization
with a
three-component mixture of $^{40}$K in the hyperfine states $%
|9/2,+5/2\rangle _{\mathrm{K}} ,|9/2,+7/2\rangle _{\mathrm{K}}$ and $%
|9/2,+9/2\rangle _{\mathrm{K}}$. The interspecies singlet and
triplet scattering lengths are nearly identical \cite{wille08},
therefore spin-exchange losses in collisions of $|3/2,+3/2\rangle
_{\mathrm{Li}}$ with $|9/2,+5/2\rangle _{\mathrm{K}}$ or
$|9/2,+7/2\rangle _{\mathrm{K}}$ are suppressed. This allows to
achieve efficient sympathetic cooling of the
lithium down to $T\simeq 10~\mathrm{\mu K}$ with $10^{5}$ atoms for both $%
^{6}$Li and $^{40}$K. For the Feshbach resonance width measurement
we
transfer the mixture into an optical dipole trap with a well depth of $%
U_{0}=360~\mu $\textrm{K} for $^{40}$K\textrm{\ }($U_{0}=160~\mu
$\textrm{K} for
$^{6}$Li) serving as an optical tweezer. The sample is transported over $22~%
\mathrm{cm}$ to a quartz cell extending from the main vacuum
chamber by moving a lens mounted on a precision linear air-bearing
translation stage. In the quartz cell we can apply homogeneous
fields ($<10~\mathrm{ppm/mm}$)\ of up to $500~\mathrm{G}$. For the
Feshbach resonance width measurement we prepare a
$|1/2,+1/2\rangle _{\mathrm{Li}}-|9/2,+9/2\rangle _{\mathrm{K}}$
mixture consisting of $4\times 10^{3}$ $^{6}$Li and $2\times
10^{4}$ $^{40}$K atoms at temperature $T\approx 21(2)~\mu
$\textrm{K}.

\begin{figure}
[ptb]
\begin{center}
\includegraphics[
trim=0.222785in 0.222654in 0.223263in 0.223138in,
natheight=4.840300in, natwidth=4.780800in, height=8.4658cm,
width=8.3516cm
]%
{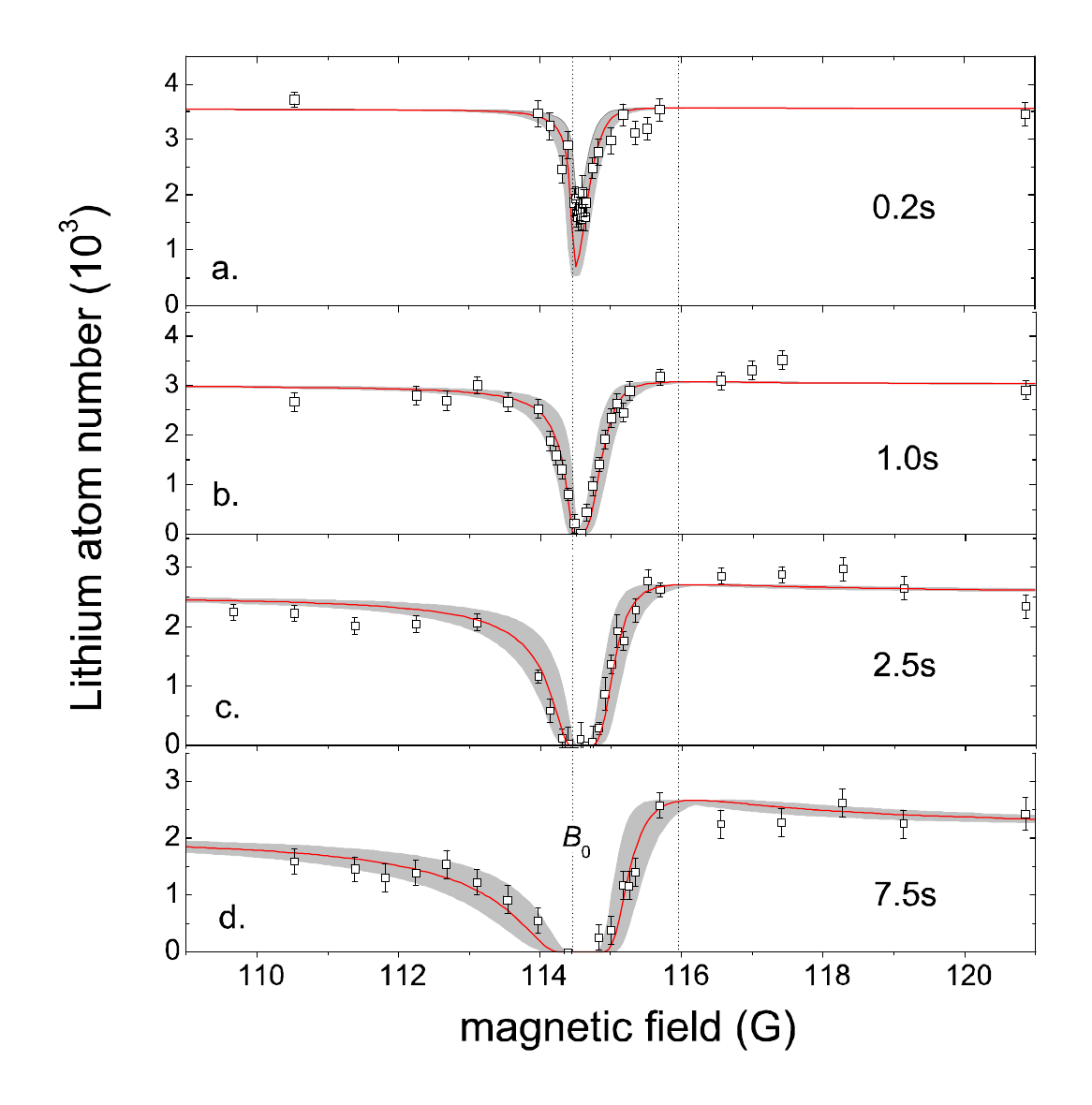}%
\caption{(Color online) Measurement of the Feshbach resonance
width. The red solid line indicates the best fit obtained for
$B_{0}=114.47(5)$ \textrm{G} and $\Delta B=1.5(5)$ \textrm{G.} The
gray shaded area indicates the combined
error in $B_{0}$ and $\Delta B$.}%
\label{fig:dataplot}%
\end{center}
\end{figure}

To observe the resonance we first ramp the field up to $\sim
107~\mathrm{G}$ where any remaining potassium spin impurities are
selectively removed by resonant light pulses. The Fano profile of
the resonance is observed by measuring the distillation rate of
the Li from the potassium-rich Li-K mixture in the optical trap as
a function of magnetic field. To initiate this process we decrease
the depth of the dipole trap in $10~\mathrm{ms}$ to
$U/U_{0}\approx 0.15$. Aside from a small spilling loss of the
$^{6}$Li this decompresses the mixture with a factor
$(U/U_{0})^{3/4}\approx 0.24$ in the
adiabatic limit and reduces the temperature accordingly by a factor $%
(U/U_{0})^{1/2}\approx 0.39$. The truncation parameter for evaporation, $%
\eta =U/k_{B}T$, drops for both species by the same amount. After
decompression the central density of the potassium is
$n_{\mathrm{K}}\approx
2\times 10^{11}\,\mathrm{cm}^{-3}$ ($n_{\mathrm{Li}}\approx 9\times 10^{9}\,%
\mathrm{cm}^{-3}$ for Li) and the temperature of the mixture is
$T=9(1)~\mu
\mathrm{K}$. As the truncation parameter of the lithium $\left( \eta _{%
\mathrm{Li}}\approx 2.7\right) $ is much smaller than that of potassium $%
\left( \eta _{\mathrm{K}}\approx 6.2\right) $ the Li
preferentially evaporates at a rate proportional to the
inter-species elastic cross section. As the lithium is the
minority component this distillation process proceeds at an
approximately constant rate. We have verified that a pure lithium
cloud experiences no rethermalization by itself. The final trap
depth $U$ was determined from the total laser power and the
measured trap
frequency for the potassium, $\omega _{r}/2\pi =1.775(6)~\mathrm{k}$\textrm{%
Hz}. In Fig.\thinspace \ref{fig:dataplot} we plot the atom number
after various holding times and as a function of magnetic field.
We analyze our data by modeling the distillation rate. Before
decompression $\left( \eta _{Li}\approx 7\right) $ we observe a
loss of $30\%$ for 1s holding time on resonance. As the
decompression reduces the density by a factor $4$\ the three-body
losses can be neglected in the decompressed trap. The
distillation of the lithium as a function of time $t$ is described by $%
N(t)=N_{0}~e^{-t/\tau _{ev}}e^{-t/\tau _{bg}}$, where
$N_{0}\approx 3\times 10^{3}$ is the initial number of lithium
atoms, $\tau _{bg}=25~\mathrm{s}$ the vacuum limited lifetime and
$\tau _{ev}^{-1}\simeq n_{\mathrm{K}}\langle \sigma \left(
k\right) \hbar k/\mu \rangle e^{-\eta _{Li}}$ the
thermally-averaged evaporation rate. Here is
\begin{equation}
\sigma (k)=4\pi \frac{a^{2}(k)}{1+k^{2}a^{2}(k)}  \label{5}
\end{equation}%
the elastic cross section with
\begin{equation}
a(k)=a_{bg}+\frac{a_{bg}\mu _{rel}\Delta B}{\hbar ^{2}k^{2}/2\mu
-\mu _{rel}\left( B-B_{0}\right) }  \label{6}
\end{equation}%
the `Doppler shifted' scattering length, with
$a_{bg}=56.6~\mathrm{a}_{0}$ at the resonance position $B_{0}$,
and $\mu _{rel}=1.57\,\mu _{B}$.

The solid lines in Fig.\thinspace\ref{fig:dataplot}\ show the best
simultaneous fit of the thermally-averaged Eq.\thinspace(\ref{5})
to the four sub-figures, accounting for $25\%$ variation in
$N_{0}$ from one day to
the next. The best fit is obtained for $B_{0}=114.47(5)$~\textrm{G} and $%
\Delta B=1.5(5)$~\textrm{G} ($R^{\ast}\approx100~\mathrm{nm}$),
where $B_{0}$
is mostly determined by the data of Fig.\thinspace\ref{fig:dataplot}a and $%
\Delta B$ by those of Fig.\thinspace\ref{fig:dataplot}d.
Uncertainties in $T$ and $n_{\mathrm{K}}$ can result in broadening
of the loss features but the difference in asymmetries between
Fig.\thinspace\ref{fig:dataplot}a-d can only originate from the
asymmetry of the elastic cross section around the resonance. The
zero crossing of $a(k)$, prominently visible in systems with
a resonantly enhanced $a_{bg}$ like $^{6}$Li \cite{zeroCrossing-Li} and $%
^{40}$K \cite{zeroCrossing-K}, remains within the noise band of
our distillation measurements because in the $^{6}$Li/$^{40}$K
system $a_{bg}$ is non-resonant
($\sigma_{bg}=1\times10^{-12}~\mathrm{cm}^{-2}$).

The investigated resonance offers good perspectives for reaching
the universal regime, for which the Fermi energy and magnetic
field have to obey: $E_{F},\mu _{rel}\left\vert B-B_{0}\right\vert
\ll \Gamma /2$, where $E_{F}\equiv \hbar ^{2}k_{F}^{2}/2\mu $ is
the characteristic relative energy of a colliding pair of atoms at
their Fermi energy. The former condition corresponds to the
condition for a broad
resonance, $k_{F}R^{\ast }\ll 1$, and is satisfied for Fermi energies $%
E_{F}\ll 5\ \mu \mathrm{K}$. The latter condition corresponds to
the condition for strong interaction, $k_{F}a\gg 1$, and is
satisfied for $\left\vert B-B_{0}\right\vert \ll 43~\mathrm{mG}$
at $E_{F}= 5\ \mu \mathrm{K}$.

In summary, we developed a model to estimate the positions and
widths of all Feshbach resonances in an ultracold gas. We selected
the optimal resonance in the $^{6}$Li/$^{40}$K system to reach the
strongly interacting regime. The experimentally observed width of
this resonance, $\Delta B=1.5(5)$~\textrm{G}, is in good agreement
with the theory and offers promising perspectives to study a
strongly interacting mass-imbalanced Fermi gas in the universal
regime using realistic experimental parameters.


We thank prof.~E.~Tiemann for stimulating discussions and S.
Whitlock for discussions on image processing. This work is part of
the research program on Quantum Gases of the Stichting voor
Fundamenteel Onderzoek der Materie (FOM), which is financially
supported by the Nederlandse Organisatie voor Wetenschappelijk
Onderzoek (NWO). We acknowledge support from the German Academic
Exchange Service (DAAD).

\end{document}